\documentclass[11pt]{article}
\usepackage{psfig}
\usepackage{multicol} 
\def\lesssim{\mathrel{\mathpalette\vereq<}}
\def\gtrsim{\mathrel{\mathpalette\vereq>}}
\makeatletter
\def\vereq#1#2{
\lower3pt\vbox{\baselineskip1.5pt \lineskip1.5pt
\ialign{$\m@th#1\hfill##\hfil$\crcr#2\crcr\sim\crcr}}}
\makeatother

\begin{document}
\begin{center}
\hfill    FERMILAB-Pub-02/060-T\\
~{} \hfill hep-ph/0204077\\

\vskip 1cm

{\large \bf Can a CPT Violating Ether Solve ALL Electron (Anti)Neutrino Puzzles?}

\vskip 1cm

Andr\' e de Gouv\^ ea

\vskip 0.5cm

{\em Theoretical Physics Division, Fermilab \\ 
 P.O. Box 500, Batavia, IL 60510, USA}

\end{center}

\vskip 1cm

\begin{abstract}
Assuming that CPT is violated in the neutrino sector seems to be a viable alternative
to sterile neutrinos when it comes to reconciling the LSND anomaly with the
remainder of the neutrino data. There are different (distinguishable) 
ways of incorporating CPT violation into the standard model, including postulating
$m\neq\bar{m}$. Here, I investigate the possibility of introducing CPT violation
via Lorentz-invariance violating effective operators (``Ether'' potentials) which
modify neutrino oscillation patterns like ordinary matter effects. 
I argue that, within a simplified two-flavor--like oscillation analysis and restricting
the analysis to the lowest order Lorentz-invariance violating oerators only, 
one cannot  solve the solar neutrino puzzle and LSND anomaly while still
respecting constraints imposed by other neutrino experiments, and comment
on whether significant improvements should be expected from a three-flavor analysis.
If one turns the picture upside down, some of the most severe
constrains on such CPT violating terms can already be obtained from the current
neutrino data, while much more severe constraints can arise from future neutrino
oscillation experiments.

\end{abstract}

\vskip 1cm

\setcounter{footnote}{0}
\setcounter{equation}{0}
\section{Introduction}


Assuming that neutrinos have a small mass and mix is by far the simplest solution 
to the two well established Solar \cite{SNP} and Atmospheric \cite{ANP} 
Neutrino Puzzles
and the more controversial LSND Anomaly \cite{LSND}. 
It is, however, well known that in order to
solve all three problems at once, one is required to add new light degrees of 
freedom to the Standard Model (SM), 
usually sterile neutrinos. The reason for this
is the fact that the Solar Neutrino Puzzle (SNP) 
requires a neutrino mass-squared difference
$\Delta m^2\lesssim 10^{-4}$~eV$^2$, the Atmospheric Neutrino 
Puzzle (ANP) requires 
$10^{-3}~{\rm eV^2}\lesssim\Delta m^2\lesssim 10^{-2}$~eV$^2$, and the LSND Anomaly
(LA) requires $\Delta m^2\gtrsim 10^{-1}$~eV$^2$. With three neutrino mass
eigenstates, one can only obtain two independent mass-squared differences.

It has recently been pointed out that there is another way to address all
neutrino puzzles with only three neutrino species: CPT violation 
\cite{Murayama_Yanagida,CPTV}.
It is easy to note that, while the SNP involves neutrinos, the LA points to 
evidence for oscillation only in antineutrinos,\footnote{Originally, the LA
manifested itself also in the neutrino channel. This effect, however, has
disappeared after more data was analyzed \cite{LSND}.} meaning that
if the $\Delta m^2$'s were different in the neutrino and antineutrino 
sectors, the three problems could be solved without the addition of light
degrees of freedom to the SM. Furthermore, CPT violation in the neutrino
sector have also been evoked as a possible solution to inconsistencies in
the neutrino data from SN1987A \cite{Murayama_Yanagida}. 

CPT is the only global space-time symmetry of the SM. Unlike 
its ``broken siblings'' (C, P, T, CP, CT, PT), however, CPT invariance is not
an accidental/optional symmetry of a quantum field theoretical system. CPT
invariance is a consequence of the fact that all microscopic phenomena 
observed to date can be perfectly described by a Lorentz invariant, local
Quantum Field Theory \cite{CPT_theorem}. 
CPT violation implies, necessarily, violation of
Lorentz invariance and/or Locality (and, perhaps, a formalism different from
Quantum Field Theory...). It has recently been stressed \cite{Greenberg}    
that allowing particle and antiparticles to have different masses seems to imply
that {\sl both} Lorentz invariance and Locality are violated.\footnote{There is
a controversy related to whether Lorentz invariance is violated under these
circumstances. I thank Gabriela Barenboim for pointing this out.}

CPT violating effects can be mimicked by neutrino interactions with
a CPT violating medium. Indeed, neutrinos and antineutrinos acquired
different {\sl effective} masses while propagating in the presence of matter,
such that large matter induced CPT violating effects can be potentially 
observed in future neutrino experiments. The authors of \cite{BBL} have
mentioned that one may be able to parameterize the CPT violating effects by
assuming the existence of a CPT violating {\sl Ether}, which modifies
the oscillation probabilities of neutrinos and antineutrinos in distinct ways.  
Certain advantages come out of such a parameterization: the theory remains 
local (only Lorentz Invariance is violated), and all well known quantum field
theoretical techniques can be safely used to explore other consequences of
the postulated CPT violation -- CPT violating effects are parametrized by
an effective Lagrangian. 
 
I argue, by performing a two-flavor analysis, that the neutrino puzzles and the LA
probably cannot be solved in a satisfactory way 
by assuming the existence of a ``CPT violating Ether.''\footnote{
The ``Ether'' is properly defined in Sec.~2.} Furthermore, it turns
out that the current neutrino data already may set
very strong limits on particular CPT violating
operators \cite{CG}, 
while potentially stronger limits may be imposed by future neutrino
experiments, such as KamLAND \cite{KamLAND} and Borexino \cite{borexino}.

In Sec.~2, the CPT violating formalism is introduced and motivated. In Sec.~3, I
discuss what are the requirements for solving the SNP and the LA within the
formalism introduced in Sec.~2, and show why it does not work
properly, at least in a two-flavor analysis. I comment briefly on whether one
should expect a three-flavor analysis to be more successful.
Sec.~4 contains a summary of the results and some discussions, 
including current and future constraints on CPT violation. 

\setcounter{equation}{0}
\section{On the Formalism}

For simplicity, I'll assume that the neutrinos are Majorana particles, and
describe it using two-component Weyl spinors. This being the case, after
electroweak symmetry breaking,
\begin{equation}
{\cal L}=i\bar{\chi}\bar{\sigma}^{\mu}\partial_{\mu}\chi-\frac{m}{2}\chi\chi-
\frac{m^*}{2}\bar{\chi}\bar{\chi}+{\cal L}_{\rm int},
\end{equation}
where $\chi$ is a two-component (left-handed) Weyl field, 
$\bar{\chi}\equiv(\chi)^{\dagger}$,
$\bar{\sigma}^{\mu}=(1,-\vec{\sigma})$, and ${\cal L}_{\rm int}$ contains
all the interaction terms.\footnote{It is worthwhile to comment in passing
that already at this level it is possible to break CPT by saying that there
are two masses $m\neq\bar{m}$. This would, however, imply in a non Hermitian Hamiltonian 
for the neutrino--antineutrino system.} 

In the spirit of \cite{CPT_formalism,CG}, we assume that possible CPT violating
effects are a consequence of unknown ultraviolet physics, and that
they are small (as required by experiments). This allows one to write CPT
violating effective operators, which are proportional to some order parameter
(v.e.v) and suppressed by powers of the new physics scale. This formalism is
particularly useful if Lorentz invariance (and CPT) is spontaneously broken 
\cite{Kostelecky_Lehnert}. As has been argued in \cite{Kostelecky_Lehnert}, this implies
that the Lagrangian is still invariant under ``observer Lorentz transformations,'' and only
``particle Lorentz invariance'' is violated. I refer readers to 
\cite{Kostelecky_Lehnert,Greenberg} for more detailed discussions.
Therefore, 
\begin{equation}
{\cal L}_{\rm CPTV}=-\sum \lambda\langle T\rangle\frac{{\cal O}_n}{\Lambda^{(n-3)}}
=-A_{\mu}\bar{\chi}\bar{\sigma}^{\mu}\chi-\sum_{n=4}^{\infty}
\lambda\langle T\rangle\frac{{\cal O}_n}{\Lambda^{(n-3)}},
\end{equation} 
where $\Lambda$ is the new physics scale, $\lambda$ are dimensionless
coupling constants, $\langle T\rangle$ is the CPT violating
order parameter and ${\cal O}_n$ are operators composed of the SM fields
with mass dimension $n$, such that 
$T{\cal O}_n$ is a Lorentz invariant operator of mass dimension $n+1$. The lowest
dimensional CPT violating operator comes at $n=3$ \cite{CPT_formalism}, and is
parametrized as $A_{\mu}\bar{\chi}\bar{\sigma}^{\mu}\chi$, where $A_{\mu}$ 
is a {\sl constant}, real four-vector with mass dimension 1. 
Note that there are no other $n=3$ CPT violating terms.
Hence forth, we ignore all other terms, which are suppressed by powers of
$1/\Lambda$, and work with the Lagrangian
\begin{equation}
{\cal L}=i\bar{\chi}\bar{\sigma}^{\mu}\partial_{\mu}\chi-
\frac{m}{2}\chi\chi-A_{\mu}\bar{\chi}\bar{\sigma}^{\mu}\chi-
\frac{m^*}{2}\bar{\chi}\bar{\chi}+{\cal L}_{\rm int}.
\label{lagrangian}
\end{equation}  

$A_{\mu}$ can be interpreted in a variety of ways. For example, it looks like
a background, classical ``electromagnetic'' field. In the case of
neutrinos, it proves useful to interpret it as a ``matter potential.'' This is
because, in the case of electron neutrinos propagating in the presence of,
say, nonrelativistic electrons, one obtains exactly 
$A_{\mu}\bar{\chi}\bar{\sigma}^{\mu}\chi$, with 
$A_{\mu}=(\sqrt{2}G_FN_e,0,0,0)$,
where $N_e$ is the electron number density. 

From Eq.~(\ref{lagrangian}), one obtains the modified Dirac equations
\begin{eqnarray}
(i\partial_{\mu}-A_{\mu})[\bar{\sigma}^{\mu}]^{\dot{\alpha}\alpha}\chi_{\alpha}-
m^*\bar{\chi}^{\dot{\alpha}}=0, \nonumber \\
(i\partial_{\mu}+A_{\mu})[\sigma^{\mu}]_
{\dot{\alpha}\alpha}\bar{\chi}^{\dot{\alpha}}-m\chi_{\alpha}=0,
\label{Dirac_eqs}
\end{eqnarray}
where $\alpha,\dot{\alpha}=1,2$ are spinor indices, and $\sigma=(1,\vec{\sigma})$. 
From Eqs.~(\ref{Dirac_eqs}), 
one can calculate the dispersion relation for both the neutrino
(positive energy state) and the antineutrino (negative energy state). The general expression
is rather cumbersome (it turn out to be a forth order polynomial dispersion relation) 
and can be found, for example, in \cite{Kostelecky_Lehnert}.

It is convenient to choose $A_{\mu}=(V,0,0,0)$\footnote{Henceforth, 
I assume that $A_{\mu}$ is time-like.} and consider the limit $V,|m|\ll E,|\vec{p}|$ 
(this is the limit I'll be strictly interested in, as will become clear in the next section)
such that \cite{CG}\footnote{see \cite{Kostelecky_Lehnert} for a discussion on the validity 
of Eq.~(\ref{disp_rel}.} 
\begin{equation}
E\simeq |\vec{p}|+\frac{m^2}{|2\vec{p}|}\pm V,
\label{disp_rel}
\end{equation}
where the plus sign applies for the neutrino and the minus sign for the 
antineutrino. The opposite sign for neutrinos and anti-neutrinos is easy to understand
if one looks at $A_{\mu}$ as a background classical vector field -- particle and
antiparticle have opposite charge.
As one can readily notice, $V$ can be interpreted as a ``potential energy'' 
and will be referred to henceforth as an ``Ether-potential.'' 
Eq.~(\ref{disp_rel}) has also been derived by several authors in the case of neutrinos
propagating in several media \cite{matter_effect_deriv}. 
In the case of ordinary matter effects, $V$ is sometimes related to an ``effective 
mass.'' We comment on this interpretation in the next section, and argue that
it can be rather misleading.

\setcounter{equation}{0}
\setcounter{footnote}{0}
\section{Addressing the Solar Puzzle plus the LSND Anomaly}

The LA can be interpreted as a measurement of $\bar{\nu}_{\mu}\rightarrow
\bar{\nu}_{e}$-conversion with $P_{\mu e}\sim 0.2\%$. 
The neutrino energies explored range (roughly) from 30~MeV to 55~MeV, 
and the neutrino travel distance is $L=30$~m. An interpretation in terms
of two-flavor neutrino oscillations indicates, after including constraints from
other experiments, that $0.1~{\rm eV^2}\lesssim\Delta m^2\lesssim 
2$~eV$^2$, while $10^{-3}\lesssim\sin^22\theta\lesssim 10^{-2}$ \cite{LSND}. 

The SNP is best interpreted by $\nu_{e}\rightarrow
\nu_{\mu,\tau}$-conversion, with $P_{ee}\lesssim 0.5$.   
The neutrino energies explored range (roughly) from 0.1~MeV to 10~MeV,
and the neutrino travel distance is (of course) one astronomical unit, and there
is information regarding the survival probability as a function of neutrino
energy. Most importantly, the presence of matter inside the Sun and the Earth affects
the oscillation pattern significantly. An interpretation in terms of
two-flavor neutrino oscillations indicates, after including constraints from
other experiments, that $10^{-9}~{\rm eV^2}\lesssim\Delta m^2\lesssim 
10^{-3}$~eV$^2$, while $0.1\lesssim\tan^2\theta\lesssim
10$\footnote{The small mixing angle solution to the SNP is currently excluded
at the 95\% confidence level.} \cite{solar_fits,FLMP}.

The question one would like to address is whether the presence of the Lorentz-invariant
violating operators 
described in the previous section can solve both the LA and the SNP. I will
restrict the discussion to two-flavor oscillations. This simplifying assumption
will render the discussion more clear, and can easily be extended to the 
three neutrino case. Note that the two-flavor case does fit trivially into 
a three flavor framework if one chooses the Ether-potentials to act only on the ``1--2''
system, and that there is no $\nu_e$ on the ``3'' mass eigenstate. The
differential equation that will lead to the oscillations in the
presence of an electron number density $N_e$ can be written as
(assuming that the neutrino energy is much larger than its mass and dropping
terms proportional to the identity matrix)
\begin{equation}
i\frac{\rm d}{{\rm d}L}
\left(\matrix{\nu_{e} \cr \nu_{x} }\right)=\left[\frac{\Delta m^2}{2E_{\nu}}
\left(\matrix{\sin^2\theta&\cos\theta\sin\theta\cr
\cos\theta\sin\theta&\cos^2\theta}\right)+
\left(\matrix{V&V_{ex}/2\cr
V_{ex}^*/2&0}\right) + \left(\matrix{\sqrt{2}G_F N_e&0\cr
0&0}\right)
\right] \left(\matrix{\nu_{e} \cr \nu_{x} }\right),
\label{de_2}
\end{equation}
where $V_{\alpha\beta}$, $\alpha,\beta=e,x$ are the Ether-potentials (as defined in the
previous section), 
$\nu_x$ is a linear combination of $\nu_{\mu}$ and $\nu_{\tau}$, and
$V\equiv V_{ee}-V_{xx}$. The 
equation for antineutrinos is identical to Eq.~(\ref{de_2}) with
$V_{\alpha\beta}\rightarrow -V_{\alpha\beta}$ and $N_e\rightarrow -N_e$.
For {\sl fixed} $V,V_{ex}$ the entire solar neutrino parameter space is spanned by
assuming $0\leq\theta\leq\pi$ after one defines the sign of $\Delta m^2$ \cite{nufact}. 
However, since we will be trying to determine $V,V_{ex}$ from neutrino data, their sign
can be adjusted such that $0\leq\theta\leq\pi/2$, as customary.\footnote{The ``dark side'' 
$\pi/4\leq\theta\leq\pi/2$ cannot be ignored due to standard matter effects 
\cite{darkside}.}

In the absence of matter effects, the oscillation probability is (these have
already been presented in \cite{CG,BPWW})
\begin{equation}
P_{ex}=P_{xe}=
\sin^22\theta_{\rm eff}\sin^2\left(\frac{\Delta_{\rm eff}}{2}L\right)
\label{pex}
\end{equation}
where
\begin{eqnarray}
&\Delta_{\rm eff}=\sqrt{(\Delta\sin2\theta+V_{ex})^2+(\Delta\cos2\theta-V)^2}, \\
&\Delta_{\rm eff}\cos2\theta_{\rm eff}=\Delta\cos2\theta-V, \\
&\Delta_{\rm eff}\sin2\theta_{\rm eff}=\Delta\sin2\theta+V_{ex}, 
\end{eqnarray}
and $\Delta\equiv \Delta m^2/2E_{\nu}$. $V_{ex}$ is assumed real 
henceforth.

For solar neutrinos, we will concentrate on solutions where the adiabatic
condition for propagation inside the Sun holds (this will always be the case 
here, as will be argued later), such that, 
for a neutrino produced close to the Sun's core,
\begin{eqnarray}
&P_{ee}=\frac{1}{2}+\frac{1}{2}\cos2\theta_M\cos2\theta_{\rm eff}, \\
&\cos2\theta_M=\frac{\Delta\cos2\theta-(\sqrt{2}G_FN_e^0+V)}
{\sqrt{(\Delta\sin2\theta+V_{ex})^2+(\Delta\cos2\theta-V-\sqrt{2}G_FN_e^0)^2}},
\end{eqnarray}
with $N_e^0$ the electron number density in the Sun's core, which translates into 
$\sqrt{2}G_F N_e^0\simeq 6\times 10^{-6}$~eV$^2/$MeV.

The next step is to search for $V,V_{ex},\Delta m^2$ and $\theta$ such that both
the LA and SNP are solved, and such that other experimental results are
not contradicted. 

First, I address the case $V_{ex}=0$. The LA requires $P_{\bar{x}\bar{e}}\sim 0.5\%$, 
assuming that 
$P_{\bar{\mu}\bar{e}}=0.5 P_{\bar{x}\bar{e}}$, such that the atmospheric
neutrino puzzle is solved by maximal $\nu_{\mu}\leftrightarrow\nu_{\tau}$
oscillations (the factor of 0.5 will play no role in the following discussions). 
Using Eq.~(\ref{pex}) for antineutrinos,
\begin{equation}
P_{\bar{x}\bar{e}}\sim \left(2.54\frac{\Delta}{\rm 1~eV^2/MeV}\sin2\theta 
\frac{L}{\rm 1~m}\right)^2, 
\end{equation}
as long as $\bar{\Delta}_{\rm eff}\lesssim 1/30$~m.\footnote{The notation 
$\bar{\Delta}_{\rm eff}$ and $\bar{\theta}_{\rm eff}$ will be used to represent
the antineutrino quantities, where $V,V_{ex}$ are replaced by $-V,-V_{ex}$
respectively.} Without any loss of generality, this will be considered as
a constraint, since the case $\bar{\Delta}_{\rm eff}\gg 1/30$ 
contradicts the Karmen data \cite{Karmen} 
and will not provide a realistic solution to the LA. 
Solving the LA implies, therefore, that 
$\Delta\sin2\theta\sim 8\times 10^{-4}$~eV$^2$/MeV for LSND-like energies
$E_{\nu}\sim$(30--55)~MeV. This in turn implies, for solar neutrinos,
$\Delta_{\rm eff}\gtrsim 3 \times 10^{-3}$~eV$^2$/MeV (remember the
largest solar neutrino energy is three times smaller than the smallest
LSND energy), such that $\Delta_{\rm eff}\gg \sqrt{2}G_F N_e$. Therefore,
matter effects inside the Sun will be very weak, and the best hope for solving the SNP
will be to obtain a very large effective mixing angle. This also explains
why the adiabatic approximation works very well.

In order to obtain $\sin^22\theta_{\rm eff}\sim 1$ one can choose
$V\sim \Delta \cos2\theta$, such that there is an MSW-like resonance  
at a typical solar neutrino energy $E_{\nu}=E^*$ between 0--10~MeV. In more detail
\begin{equation}
\sin^2 2\theta_{\rm eff}=\left[
1+\left(\frac{\delta E}{E^*\tan2\theta}\right)^2\right]^{-1}, 
\end{equation}
where $\delta E\equiv E_{\nu}-E^*$ and $E^*$ is the resonant solar neutrino
energy. The energy dependence of the solar neutrino data requires, 
very generously, that, for $\delta E\sim E^*$, $\sin^22\theta_{\rm eff}\gtrsim 
0.6$.\footnote{Under these circumstances, 
$P_{ee}$ is constrained to be between 0.5 and 0.7 for the entire solar 
neutrino energy range. This provides a rather poor fit to the data, but is the 
best one can hope to achieve.} This translates into $\tan^22\theta>3/2$. It turns
out that the CHOOZ bound \cite{CHOOZ} 
on $P_{\bar{e}\bar{e}}$ forbids such a choice.

At CHOOZ, $\bar{\Delta}_{\rm eff}\gtrsim 3\times 10^{-3}$~eV$^2$/MeV (remember
that typical reactor antineutrino energies are of the order of typical solar
neutrinos energies), such that $1-P_{\bar{e}\bar{e}}
\sim 1/2\sin^22\bar{\theta}_{\rm eff}$ at the CHOOZ experiment ($L_{\rm CHOOZ}
\sim 1000$~m). Assuming that there is a resonance in the solar neutrino sector
at some $E_{\nu}=E^*$,
\begin{equation}
\sin^22\bar{\theta}_{\rm eff}=
\frac{\tan^22\theta}{\tan^22\theta+(1+E_{\bar{\nu}}/E^*)^2},
\end{equation}   
and the bound $\sin^22\theta_{\rm eff}<0.1$ from CHOOZ translates
into $\tan^2\theta< 4/9$, in disagreement with the requirements from the the
SNP discussed above.

Next, consider $V=0$ but $V_{ex}\neq 0$. As before, the LA will constrain
$(\bar{\Delta}_{\rm eff}\sin2\bar{\theta}_{\rm eff})^2=
(\Delta \sin2\theta-V_{ex})^2\simeq (8\times 10^{-4})^2~\rm eV^4/MeV^2$. In 
the case $|V_{ex}|<\Delta \sin2\theta$, the survival probability will
be very similar to the case of no Lorentz invariance violation 
($\Delta_{\rm solar}\gg \Delta_{\rm LSND}$), and is therefore uninteresting.

The opposite hypothesis, $|V_{ex}|\gg \Delta \sin2\theta$ for typical LSND
energies invites further investigation. In this case, the LA implies 
$|V_{ex}|\sim 8\times 10^{-4}$~eV$^2$/MeV. If $|V_{ex}|\gg \Delta \sin2\theta$ 
also for typical solar neutrino energies, $\sin^22\theta_{\rm eff}\sim 1$
for solar neutrinos, which provides a reasonable solution to the SNP. 
This possibility is, unfortunately, immediately killed by the CHOOZ bound, since
$\sin^22\bar{\theta}_{\rm eff}\sim 1$ for typical reactor antineutrino energies.

Another possibility, which can be quickly discarded, is the case $V_{ex}+\Delta 
\sin2\theta=0$ and $\Delta\cos2\theta$ very small for typical solar neutrino
energies, such that $\Delta_{\rm eff}\sim \sqrt{2}G_FN_e^0$. In this case, 
$\bar{\Delta}_{\rm eff}\sim V_{ex}$ is large, such that the CHOOZ bound 
would be grossly violated, namely, $1-P_{\bar{e}\bar{e}}\sim 1/2$ at CHOOZ.

Finally, there is the possibility of evoking an ``antiresonance'' at CHOOZ,
namely $\Delta \sin2\theta-V_{ex}=0$ for some reactor antineutrino
energies $E_{\bar{\nu}}=\bar{E}^*$. In this 
case, one may hope to obtain maximal effective mixing in the neutrino sector
and solve the SNP. In more detail
\begin{equation}
\sin^22\bar{\theta}_{\rm eff}=\frac{(\delta E/\bar{E}^*)^2}
{(\delta E/\bar{E}^*)^2 + \rm cotan^22\theta},
\end{equation}   
where $\delta E\equiv E_{\bar{\nu}}-\bar{E}^*$. Note that 
away from $\bar{E}^*$, the CHOOZ bound still presents a limit for $\tan^2\theta$.
For example, requiring $1-P_{\bar{e}\bar{e}}<0.5$ for
$\delta E/\bar{E}^*=1/2$ implies $\tan^22\theta<4/9$. 

Under these circumstances, the solar effective angle is
\begin{equation}
\sin^22\theta_{\rm eff}=\frac{(1+E_{\nu}/\bar{E}^*)^2}{ (1+E_{\nu}/
\bar{E}^*)^2 + 
\rm cotan^22\theta},
\end{equation}
which, in the limit $E_{\nu}\ll \bar{E}^*$ implies $P_{ee}\gtrsim 0.85$,
in disagreement with the solar data.

It worthwhile to comment at this point that it should not come as a surprise
that no remotely appropriate fit to all neutrino data can be obtained with
either $\Delta m^2,\theta,$ and $V$, or $\Delta m^2,\theta,$ and $V_{ex}$
as free parameters: the fit is severely over constrained. It remains, however, to
check whether the addition of a fourth free parameter can help significantly.
There is some room for optimism, since nonzero $V$ and $V_{ex}$ seem to play
significantly different roles in the previous discussions. For example,
one may envision a large $V_{ex}$ which can take care of the CHOOZ bound,
combined with a small enough $\theta$ and a large enough $\Delta m^2$ to
solve the LA. The SNP may be solved by appropriately choosing $V$ in order
to enforce a large effective solar mixing angle, and so on.

\begin{figure}[ht]
\centerline{
\psfig{file=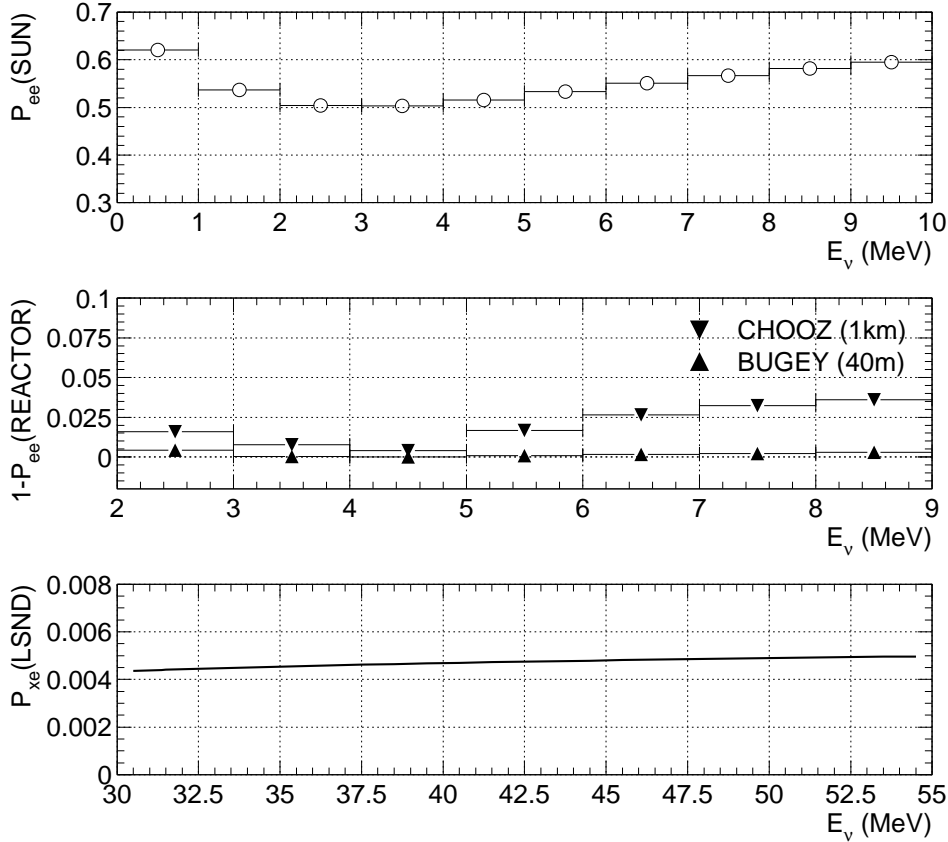,width=.75\columnwidth}}
\caption{TOP -- $P_{ee}$ for solar neutrinos, CENTER -- $1-P_{\bar{e}
\bar{e}}$ for reactor antineutrinos (at both the CHOOZ Bugey experiments) 
and BOTTOM -- $P_{\bar{x}\bar{e}}\sim 2\times P_{\bar{\mu}\bar{e}}$ at
the LSND experiment, as a function of the (anti)neutrino energy, for
$\Delta m^2=0.01$~eV$^2$, $\cos 2\theta=0.6$, $V=0.001$~eV$^2$/MeV, and 
$V_{ex}=0.001$~eV$^2$/MeV (see text for details).}
\label{fig:best}
\end{figure}

An optimal solution was numerically searched by performing a ``straw-man
fit'' to the CHOOZ, solar and LSND data. Explicitly, 
this was done by requiring that
$P_{ee}=0.4 \pm 0.1$ in ten 1~MeV-wide solar neutrino energy bins 
(from 0 to 10~MeV), $1-P_{\bar{e}
\bar{e}}=0\pm0.04$ in seven 1~MeV-wide reactor antineutrino energy (from 2 to 9~MeV) 
bins at CHOOZ, and $P_{\bar{\mu}\bar{e}}=0.25\%\pm 0.08\%$ in five 5~MeV-wide 
antineutrino energy bins at LSND (from 30 to 55~MeV). A more detailed fit is 
beyond the scope of this discussion, but it should be noted that the 
constraints imposed here are rather mild, such that a realistic fit will 
probably yield more severe constraints on a Lorentz-invariant violating 
solution to both the SNP 
and LA. The oscillation probabilities obtained at the solar, reactor, and   
LSND (anti)neutrino energy ranges are depicted in Fig.~\ref{fig:best}, for
one ``best fit point:'' 
$\Delta m^2=0.01$~eV$^2$, $\cos 2\theta=0.6$, $V=0.001$~eV$^2$/MeV, and 
$V_{ex}=0.001$~eV$^2$/MeV. Other points which also provide a reasonable fit
are very similar, and characterized by $0.0007\lesssim V_{ex}\lesssim 0.0012$,
$0.0001\lesssim V\lesssim 0.001$, $0.5 \lesssim \cos 2\theta\lesssim 0.7$,
$0.005\lesssim \Delta m^2 \lesssim 0.015$~eV$^2$. 

A few comments are in order. First of all, the result obtained is not
altogether inconsistent with the data. As a matter of fact, the result obtained
is in good agreement with the reactor constraints, and the LSND data. The
Karmen constraints are easily satisfied, given that $\bar{\Delta}_{eff}\ll 
1/L$ at both LSND and Karmen, such that oscillation effects at the Karmen
site are suppressed with respect to effects at the LSND experiment by 
$(30/18)^2$. (remember that $P_{\bar{\mu}\bar{e}}\simeq 0.5\times
P_{\bar{x}\bar{e}}$). The biggest concern rests in the solution to the SNP.
The fact that $P_{ee}>0.5$ in the entire energy range (the reason for
this, namely the fact that $\Delta_{\rm eff}\gg \sqrt{2}G_FN_e^0$ has 
already been addressed in detail), implies that
a good fit to the solar data cannot be obtained. For example, the SNO and
SuperKamiokande data alone rule out this model at more than the two sigma 
level.\footnote{See \cite{FLMP} for a detailed discussion. The best possible
fit (already excluded at more than the 2 sigma level) 
would be obtained if the Standard Solar Model prediction for 
the $^8$B solar neutrino flux \cite{SSM} was off by a factor of 2.} Nonetheless,
the energy dependency of $P_{ee}$ obtained for the Lorentz-invariant violating
model considered here
is not too different from that obtained for regular two-neutrino oscillations
and $\Delta m^2_{\rm solar}> \rm few \times 10^{-4}$~eV$^2$.

Next, I address the atmospheric neutrino data. For atmospheric (anti)neutrinos,
which travel at least 10~km before reaching the, say, SuperKamiokande detector,
$L\gg 1/\Delta_{\rm eff},1/\bar{\Delta}_{\rm eff}$ assuming that the values of the
four parameters $\theta,\Delta m^2,V,V_{ex}$ provide a proper fit to the SNP and
the LA. This means that $P_{ee}\simeq 1- 1/2 \sin^22\theta_{\rm eff}$. Furthermore,
because $\Delta_{\rm atm}\lesssim 10^{-5}~{\rm eV^2/MeV}\ll V,V_{ex}$, 
$\sin^2\theta_{\rm eff}\simeq (V_{ex})^2/((V_{ex})^2+(V)^2)$. Given the range
of parameters delimited above, one obtains
\begin{equation}
0.75\lesssim P_{ee}\simeq P_{\bar{e}\bar{e}}\lesssim 0.5,
\end{equation}    
for all $E_{\nu}$ and all $L$. Such a possibility is certainly not a good fit
to the atmospheric neutrino data, which clearly prefers $P_{ee}\simeq P_{\bar{e}\bar{e}}
\simeq 1$. In order to be able to make a definitive statement, however,
a full three neutrino analysis of the atmospheric data including the Ether-potentials
is required. Nonetheless, there is a hint that the Lorentz-invariant violating hypothesis
discussed here is strongly disfavored if one considers the current constraint
on $|U_{e3}|^2$ from the ANP \cite{three_nu_atm}. A realistic analysis may
provide even stronger constraints, given that the value of $P_{ee},P_{\bar{e}\bar{e}}$
above are energy independent, and should also modify the measured
$\nu_{\mu}$-flux distributions significantly (given that $1-P_{ee}\sim 0.5 P_{\mu e}$).   

Finally, the most stringent constraints come from old $\bar{\nu}_{\mu}\rightarrow\bar{\nu}_e$
searches \cite{new_ref}.\footnote{I thank Alessandro Strumia for pointing this out. The fact that 
severe constraints should be obtained from these experiments was first alluded to in 
\cite{Strumia}.}
Similar to atmospheric neutrinos, the neutrinos in these experiments
are high energy and the baseline is long, such that 
$0.25\lesssim P_{xe}\sim P_{\bar{x}\bar{e}}\lesssim 0.5$. The result of \cite{new_ref}, namely
$P_{\bar{\mu}\bar{e}}<6.5\times 10^{-3}$, is in gross disagreement with the ``best fit'' 
expectations quoted above.  

In summary, it is fair to say that the presence of Ether-potentials (as defined in the previous
section) cannot 
accommodate the current neutrino data, at least when only
the $\nu_e\leftrightarrow \nu_x$-sector is affected ($\nu_x$ is a linear combination of
$\nu_{\mu},\nu_{\tau}$). 
While a good fit to the LA and the reactor data
can be obtained, only a marginal solution to the SNP exists, and there are further 
constraints from the atmospheric data and older searches for $\bar{\nu}_{\mu}$ oscillations. 
However, the fact that one cannot fit all the data
should not come as a surprise. The most important reason for this expectation is
that the Lorentz-invariant violating 
term serve as an effective potential, and not an effective mass. This
implies that a) the standard $L/E$ behavior of the survival probabilities will 
be severely modified, b) the effective mixing angle will vary substantially with energy
and c) standard matter effects are altered. Indeed, the
fact that the SNP and the LA can be marginally
reconciled without violating the CHOOZ bound
may come as a bigger surprise (at least to me)!

An important question is whether a full three neutrino fit could be more successful. 
Such a fit would contain eleven free parameters (two mass-squared differences, three 
mixing angles, one complex Dirac phase and five Ether-potentials) and is certainly beyond 
the scope of this discussion. While there are enough new free parameters (which should
render performing such a fit very challenging), it is worthwhile to comment
that many features of the atmospheric data, such as
the peculiar $L/E$ of the $\nu_{\mu}$ flux should prove to be a serious challenge to the
oscillations in the presence of the Lorentz-invariance violating operators considered here, 
indicating that such a fit will encounter 
challenges which are not too dissimilar from the ones faced here. I believe that obtaining
a successful fit to all the data with a three-neutrino is rather improbable, but,
perhaps, not impossible. 

\setcounter{equation}{0}
\section{Summary and Discussions}

CPT violation implies, necessarily, violation of Lorentz invariance and/or Locality.
Under certain circumstances, CPT violation requires a reinterpretation of the 
Quantum Field Theoretical formalism which is used to describe, extremely successfully, 
all short distance physics to date.

It has been pointed out that if the neutrinos and antineutrinos had different
masses (CPT violation), all neutrino puzzles could be solved without the addition of   
extra light degrees of freedom to the standard model. This hypothesis seems
to violate
both Lorentz invariance and Locality, and renders a Quantum Field theoretical
description of the neutrinos (including loop-effects, time-evolution, etc) rather
cumbersome.   

Here, I have considered a formalism for CPT violation \cite{CG,CPT_formalism} 
which can be treated
self consistently within Quantum Field Theory (it look like ``standard'' effective
field theory) by adding a few effective Lorentz invariant-violating operator 
(Ether-potentials). It turns out, however, that one cannot obtain a set of parameters
which will satisfy the solar neutrino data, the atmospheric neutrino data, the 
LSND data, and the reactor data at the same time, at least within a two-flavor
analysis. I tried to argued that, 
contrary to previous claims \cite{BBL}, one should at least suspect that the existence of
Ether-potentials is insufficient to accommodate all the neutrino data. The most interesting
feature discussed here is, perhaps, the fact that a good fit can almost be obtained
if both diagonal and, more importantly, off-diagonal Ether-potentials are present.
One of the big barriers one is forced to face is the fact that, for solar neutrinos,
the electron neutrino survival probability
is always greater than one half, in contradiction with
the current solar neutrino data ({\it e.g.,}\/ SuperKamiokande and SNO). If it turns
out that the solar data definitively {\sl requires} $P_{ee}<0.5$ (at a very high
confidence level) for a finite energy range, the ``Ether solution'' would be unambiguously 
ruled out. This may very well be achieve after the new SNO results on the neutral
current cross-section and the day-night effect are released.

Constraints on $V$ and $V_{ex}$ (as defined in the previous section) are, of
course, possible to obtain outside the realm of neutrino oscillations. The authors
of \cite{MP} already point out that, through loops, one can severely constraint
$V_{ee}$.\footnote{\cite{MP} considers a ``seesaw''-like origin for the neutrino masses
and the CPT violating operators, such that some ``translation'' to the
formalism used here is required in order to read off the bound on $V_{ee}$.} Since
oscillation signatures only depend on $V_{ee}-V_{\mu\mu}$, this strong bound
is irrelevant if one allows $V_{ee}\ll V_{\mu\mu}$. On the other hand, 
if the Ether-potentials were to be written in $SU(2)_L\times U(1)_Y$ invariant forms, 
very severe tree-level constraints from the charged lepton sector would probably render the
model useless as far as neutrino oscillation phenomenology is concerned.

Constraints on the off-diagonal $V_{ex}$ and on the second generation $V_{\mu\mu}$ 
are expected to be
much weaker, and the best constraints will come from the neutrino oscillations
experiments themselves. For example, from the discussions in the previous section,
the LA already constraints $|V|,|V_{ex}|\lesssim 10^{-3}~\rm eV^2/MeV=10^{-18}~GeV$. 
Future neutrino experiments including KamLAND (\cite{Bahcall_CPT} for a detailed
discussion), Borexino and long baseline neutrino experiments (for a detailed discussion,
see \cite{BPWW}) can push the bounds on $V_{\alpha\beta}$ by many orders
of magnitude. For example, if it turns out that strong solar matter effects are required
to solve the SNP, one would be force to constraint $|V|,|V_{ex}|\lesssim 
10^{-5}~\rm eV^2/MeV=10^{-20}$~GeV, while if Borexino observes seasonal 
variations, their data will be consistent with 
$\Delta m^2/2E_{\nu}\lesssim 10^{-9}~\rm eV^2/MeV=10^{-24}$~GeV \cite{seasonal}, 
implying $|V|,|V_{ex}|\lesssim 10^{-24}$~GeV.  

It is important to note that other Lorentz invariant violating effects which differ from
the ones considered here also lead to modified neutrino oscillation probability.\footnote{It
is possible to obtain, for example, a neutrino oscillation phase proportional to 
$L\times E$ \cite{CG,VEP}
instead of the mass induced $L/E$ behavior or the $L$ behavior discussed here.}
For a constraints imposed on such terms by the atmospheric neutrino data see \cite{Fogli:1999}. 
The results presented here apply only to the scenario when only the Ether-potentials are 
added to the Standard Model, and it is certainly not clear whether a complicated combination 
of several different effects will not lead to a better fit to the data. It does seem, in
my opinion, unlikely.
 
\section*{Note Added}

After the completion of this work, new data was presented by the SNO collaboration.
It strengthens significantly the statement that $P_{ee}<0.5$ for $^8B$ neutrinos (at more
than the three sigma level) and renders the possibility of obtaining a good fit with the 
Ether-potentials attempted here virtually impossible, as hinted to in Sec.~4. The SMA solution
to the solar neutrino puzzle, which was ignored in the discussions contained here, is 
ruled out by the current solar data at around the five sigma level.

\section*{Acknowledgements}

I am happy to thank Gabriela Barenboim, John Beacom, and Boris Kayser 
for suggestions and comments on the manuscript. I also thank Alessandro Strumia
for very useful comments. This work was supported
by Fermilab, which is operated by the URA under DOE contract 
No. DE-AC02-76CH03000. 



\end{document}